\documentclass[osajnl,twocolumn,showpacs,superscriptaddress,11pt]{revtex4-1} 
\usepackage{amsmath,amssymb,graphicx,longtable,psfrag,subfigure,multirow}
\begin{document}

\title{High sensitivity fiber optic angular displacement sensor and its application for detection of ultrasound}

\author{Jo{\~a}o M. S. Sakamoto}\email{Corresponding author: sakamoto@ieav.cta.br}
\affiliation{Division of Photonics, Instituto de Estudos Avan{\c c}ados, Trevo Cel. Av. Jos{\'e} A. A. do Amarante n$^o$ 1, S{\~a}o Jos{\'e} dos Campos, SP 12228-001, Brazil}

\author{Cl{\'a}udio Kitano}
\affiliation{Department of Electric Engineering, Universidade Estadual Paulista, Campus III, Ilha Solteira, SP 15385-000, Brazil}

\author{Gefeson M. Pacheco}
\affiliation{Department of Microwave and Optoelectronics, Instituto Tecnol{\'o}gico de Aeron{\'a}utica, Pra{\c c}a Mal. Eduardo Gomes n$^o$ 50, S{\~a}o Jos{\'e} dos Campos, SP 12228-900, Brazil}

\author{Bernhard R. Tittmann}
\affiliation{Department of Engineering Sciences and Mechanics, Pennsylvania State University, 212 Earth \& Engr. Sciences Building, University Park, PA 16802-6812, USA}

\begin{abstract}
In this paper, we report the development of an intensity modulated fiber optic sensor for angular displacement measurement. This sensor was designed to present high sensitivity, linear response, wide bandwidth and, furthermore, to be simple and low cost. The sensor comprises two optical fibers, a positive lens, a reflective surface, an optical source, and a photodetector. A mathematical model was developed to determine and simulate the static characteristic curve of the sensor and to compare different sensor configurations regarding the core radii of the optical fibers. The simulation results showed that the sensor configurations tested are highly sensitive to small angle variation (in the range of microradians) with nonlinearity less than or equal to 1\%. The normalized sensitivity ranges from $(0.25\times V_{\rm max})$ to $(2.40\times V_{\rm max})$ mV/$\mu$rad (where $V_{\rm max}$ is the peak voltage of the static characteristic curve) and the linear range, from 194 to 1840 $\mu$rad. The unnormalized sensitivity for a reflective surface with reflectivity of 100\% was measured as 7.7 mV/$\mu$rad. The simulations were compared with experimental results to validate the mathematical model and to define the most suitable configuration for ultrasonic detection. The sensor was tested on the characterization of a piezoelectric transducer and as part of a laser ultrasonics setup. The velocity of the longitudinal, shear, and surface waves were measured on aluminum samples as 6.43 mm/$\mu$s, 3.17 mm/$\mu$s, 2.96 mm/$\mu$s, respectively, with an error smaller than 1.3\%. The sensor proved to be suitable to detect ultrasonic waves and to perform time-of-flight measurements and nondestructive inspection, being an alternative to the piezoelectric or the interferometric detectors.
\end{abstract}


\maketitle 

\section{Introduction}
Laser ultrasonics is an all-optical nondestructive inspection technique which employs a Q-switched laser (a pulsed laser) to generate an ultrasonic wave over the surface (or bulk) of an inspected sample, and an optical sensor to detect the ultrasonic wave (after crossing the region under analysis) and to provide an electric signal containing the required information. Inasmuch as the aforementioned technique employs an all-optical setup, it becomes attractive, since it is couplant free, non-contact and remote from the inspected sample \cite{scruby1990a}. The most common method to generate and detect ultrasound, in order to perform nondestructive inspection, employs the piezoelectric transducer. Besides being a contact transducer, one can cite several disadvantages as the requirement of a coupling medium, high temperatures (over the Curie point of the piezoelectric material) are not supported, narrow bandwidth, low spatial resolution, loading of the sample surface, and it is hard to use in complex or curved geometries \cite{scruby1990a, murfin2000, monchalin2002, sorazu2003}. To overcome the drawbacks of the piezoelectric transducer, optical detectors as interferometric (e.g., Mach-Zehnder, Michelson, Fabry-Perot, Photorefractive) or non-interferometric configurations (e.g., knife-edge, fiber optic based sensors, microring sensors) have being applied and developed \cite{monchalin1986, murfin2000, monchalin2004, bramhavar2009, beard1997, ling2011}. Among these, for industrial applications, the interferometric detectors stands out due to their high sensitivity and their ability to detect on rough surfaces \cite{dewhurst1999}. However, its cost is high as is the complexity of the setup \cite{murfin2000, perret2011}. An alternative to the interferometer is the intensity modulated fiber optic sensor, which can have a high sensitivity, can be low cost and simple.

In the literature, there are a number of intensity modulated fiber optic sensors, where can be highlighted the linear displacement sensors which basic configuration comprises two fibers (emitting and receiving fibers) and a reflective surface. This sensor can measure linear displacement between the fibers and the reflective surface, relating it to the optical intensity reflected by the surface and coupled into the receiving fiber. Different configurations can be found in the literature with the emitting and receiving fibers positioned parallel to each other or in angle; instead of using only two fibers, a bundle of emitting fibers and/or a bundle of receiving fibers can be used to increase the sensitivity \cite{menadier1967, he1991, ko1995, shimamoto1996, bergougnoux1998, faria1998, zheng1999, bucaro2001, buchade2006, buchade2007, patil2011, perret2011, moro2011}. A drawback of this approach is the inability to distinguish linear displacement from angular displacement, decreasing the accuracy of the sensor \cite{sagrario1998}. 
Another version, based on the same configuration but modified to measure angular displacement, comprises an additional positive lens \cite{sagrario1998, bois1989, wu1995, khiat2010}. The configurations proposed by these authors require fine adjustment and complex mounting. In addition, not much attention is paid to the lens position; it is used essentially to focus the light over the reflective surface. According to \cite{wang1997}, an improvement in the amount of light collected can be accomplished collimating the light beam through the lens. The sensor proposed in \cite{wang1997} uses a graded-index (GRIN) lens to collimate the light and the fibers are placed in angle, facing the reflective surface. However, the sensor was constructed to measure linear displacement and the angular displacement is treated as a secondary effect. The sensor proposed in \cite{feldmann2005} uses an integrated lens to collimate the light and the fibers are also placed in angle. According to the authors, the sensor can measure distance and it is applied as a microphone. The complexity and cost, however, are high, since the manufacturing requires lithographic steps. 

In this work, it was developed an intensity modulated fiber optic sensor for angular displacement measurement. This sensor was designed to present high sensitivity, linear response, wide bandwidth and, furthermore, to be simple and low cost. The configuration proposed, as some angular displacement sensors, comprises two fibers, a reflective surface, and a positive lens. The difference arise in the arrangement of the fibers, placed parallel to each other, combined with the arrangement of the lens, placed in order to collimate the light beam. This configuration allows unequivocal measurement since it brings very low sensitivity to linear displacement and, on the other hand, very high sensitivity to angular displacement \cite{sakamoto2010}. Besides, as the laser beam is collimated, the axial position of the reflective surface can be chosen according to the requirements of the measurement (e.g., long stand-off distances can be set) and a fine positioning is not necessary. The mathematical model of this sensor was presented in order to determine its static characteristic curve according to geometrical parameters. Computational simulation (based on the mathematical model) and experimental data were acquired for nine different sensor's configuration (regarding the core radii of the fibers), validating the mathematical model and allowing a comparison between the configurations. Finally, the sensor proposed was applied to detect ultrasound directly from the surface of a piezoelectric transducer and furthermore as the optical detector of a laser ultrasonics setup, showing to be able to detect longitudinal, shear, and surface ultrasonic waves.

Besides the laser ultrasonics application (the motivation of this work), typical applications of this sensor could be reflective surface angle measurements itself, as a microphone or hydrophone (with a suitable diaphragm or membrane), for acoustic emission sensing, characterization of piezoelectric actuators, pulse-echo monitoring, measurement of ultrasonic velocity and attenuation, and as part of an atomic force microscope (AFM) \cite{wu1995, murfin2000, perret2011}.

\section{Principle of operation}
\begin{figure}[ht]
  \centering  
  \includegraphics[width=8cm]{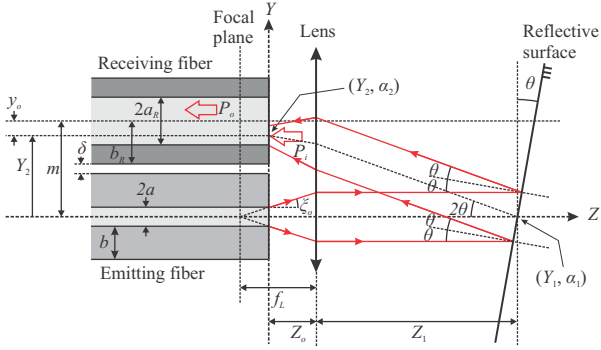}
  \caption{(Color online) Sensor head.}
  \label{fig:SH}
\end{figure}

As stated before, the two fibers are positioned and fixed parallel to each other, with their tips aligned. The positive lens is positioned in front of the fibers and the reflective surface, in front of the lens, as shown in Fig.~\ref{fig:SH}.
The lens plays two roles: in one direction it collimates the light from the emitting fiber; in the opposite direction it focuses back the light reflected by the reflective surface. The principle of operation of the sensor is based on the lens effect, which converts the angular variation of the collimated beam (caused by the reflective surface), $\theta$, in variation of the beam spot's position at the tip of the receiving fiber, $Y_2$. The receiving fiber, besides collecting the light, works as a knife-edge and the spot position $Y_2$ determines the amount of light coupled to it. 

Based on this principle, a static characteristic curve can be acquired determining the power transfer coefficient, $\eta$ (defined as the ratio between the optical power coupled into the receiving fiber, $P_o$, and the total optical power incident at the receiving fiber surface, $P_i$), for each $\theta$ value. This curve presents two linear regions which can be used for dynamic measurements. To accomplish a dynamic operation, the reflective surface angle must be adjusted in order to set the operation point on the center of the linear region. In the case of ultrasound detection, the reflective surface should be replaced by the sample under analysis and the angle should be adjusted properly. The ultrasonic wave that reaches the surface of the sample creates a mechanical disturbance, which can be seen as a peak and a valley with an inclined region between then. Therefore, the sensor is able to detect the ultrasonic wave since this inclined region is function of the angle $\theta$.

The sensor's circuitry is based on a transimpedance scheme designed to acquire signals from DC up to tens of megahertz. Thus, the sensor provides an output voltage signal in the time domain directly, with no additional circuits for demodulation as occurs in the interferometer configurations. Also, this sensor does not require a lock-in amplifier in dynamic operation (since the ultrasonic frequencies are much higher than the environmental vibration) or additional optical components.

In the next section, the mathematical model is presented in order to determine the power transfer coefficient, $\eta$, as function of the reflective surface angle, $\theta$.

\section{Mathematical model}
In Fig.~\ref{fig:SH}, the following parameters are defined: $a$ is the emitting fiber core radius, $b$ is the emitting fiber cladding radius, $a_R$ is the receiving fiber core radius, $b_R$ is the receiving fiber cladding radius, $\delta$ is the gap separation between the two fibers, $Z_o$ is the distance between the lens and the fibers, $Z_1$ is the distance between the lens and the reflective surface, and $f_L$ is the lens focal distance. The $XYZ$ axis is fixed, with origin on the center of the emitting fiber and $Z$ along the emitting fiber axis.

An optical source provides the light that is conveyed by the emitting fiber and the light emerges from it in a conical shape limited by the critical angle $\xi_o$, reaching the lens. This angle is given by:
\begin{equation}
\xi_o = {\sin^{-1} (NA/n)},
\label{eq:xio}
\end{equation}
where $NA$ is the emitting fiber numerical aperture and $n$ is the index of refraction of the medium around the fibers (air in this case). The distance $Z_o$ is set to obtain a collimated beam after the lens, given by:
\begin{equation}
Z_o = f_L - a/{\tan \xi_o}.
\label{eq:zo}
\end{equation}
The collimated beam reaches the reflective surface with incidence angle $\theta$ relative to its normal, it is reflected with the same angle, and the total deviation angle is equal to $2\theta$. The reflected beam stays collimated, traveling in the opposite direction and impinges the lens again. Then, the lens focuses the beam and the optical spot reaches the receiving fiber core.

To determine the static characteristic curve equation, $\eta(\theta)$, an expression relating $\theta$ and the spot center position, $Y_2$, is found analyzing the chief ray of the reflected beam. Knowing the chief ray position and angle $(Y_1,\alpha_1)$ at the reflective surface and using an ABCD matrix, it is possible to determine the position and angle $(Y_2, \alpha_2)$ at the receiving fiber plane ($Z=0$). Regarding the chosen axis shown in Fig.~\ref{fig:SH}, $Y_1 = 0$ and $\alpha_1 = 2 \theta$. The angle $\theta$ is small enough to use a paraxial approximation and the following ABCD matrix \cite{siegman1971}:
\begin{equation}
\left[ \begin{array}{c} Y_2 \\ \alpha_2 \end{array} \right] = \left[\begin{array}{lr} 1 & Z_o\\ 0 & 1 \end{array} \right]
\left[\begin{array}{cr} 1 & 0\\ \frac{-1}{f_L} & 1 \end{array} \right]
\left[\begin{array}{lr} 1 & Z_1\\ 0 & 1 \end{array} \right]
\left[ \begin{array}{c} 0 \\ 2\theta \end{array} \right] .
\label{eq:ABCD}
\end{equation}

The angle $\alpha_2$ is not necessary in the remaining calculation since the optical spot radius is regarded as constant and circular, due to the paraxial approximation. 
Thus, evaluation of Eq. \ref{eq:ABCD} yields:
\begin{equation}
Y_2(\theta)=K\theta,
\label{eq:Y2}
\end{equation}
where $K = 2(Z_o+Z_1-Z_1Z_o/f_L)$. 

\begin{figure}[ht]
  \centering  
  \includegraphics[width=8cm]{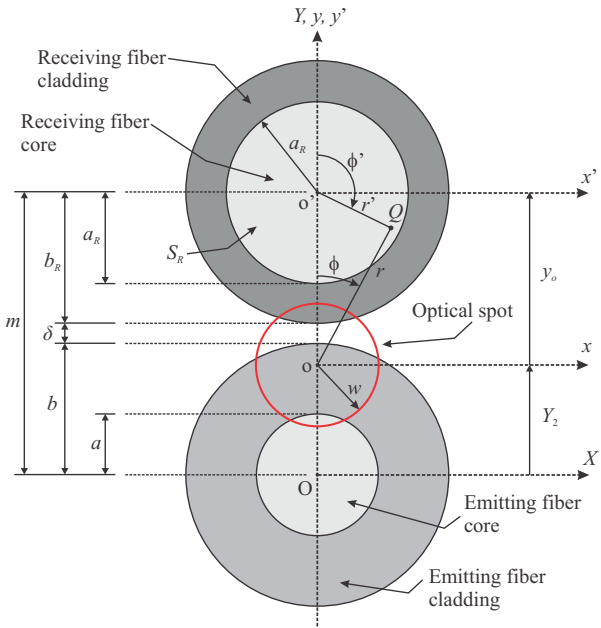}
  \caption{(Color online) Front view of the sensor head and the optical spot.}
  \label{fig:Integ2}
\end{figure}

The origin of the coordinate system $xyz$ is located at the spot center, as shown in Fig.~\ref{fig:Integ2}.
The intensity profile of the optical spot can be regarded as a gaussian, and can be written in cylindrical coordinates in terms of $P_i$ as \cite{siegman1971a}:
\begin{equation}
I(r) = \frac{2 P_i}{\pi w^2} \exp\left(\frac{-2 r^2}{w^2}\right)
\label{eq:Ir2}
\end{equation}
where $r$ is the radial coordinate related to the $xy$ axis, and $w$ is the optical spot radius. The beam divergence can be disregarded since the distance $Z_1$ is kept lower than half of the Rayleigh range, $z_R$, given by \cite{siegman1971}: $z_R= \pi w_o^2/\lambda$, where $w_o$ is the beam waist and $\lambda$ is the wavelength of the optical source. In this way, the optical spot radius can be regarded as approximately the same size of the emitting fiber core radius, i.e., $w \approx a$.

The optical power coupled into the receiving fiber, $P_o$, can be evaluated integrating Eq. \ref{eq:Ir2} over the receiving fiber core area, $S_R$, as:
\begin{equation}
P_o = \Gamma \int\!\!\!{\int_{S_R} {I(r) {\rm d}S_R}},
\label{eq:Po}
\end{equation}
where ${\rm d} S_R = r {\rm d}\phi {\rm d}r$ is the differential element of area and the constant $\Gamma$ ($0 \leq \Gamma \leq 1$) accounts for the transmission loss.
The receiving fiber, has at its center, the origin of the coordinate system $x'y'z'$, corresponding to the radial coordinate $r'$, as shown in Fig.~\ref{fig:Integ2}. The coordinates of an arbitrary point $Q$, are defined in cylindrical coordinates as ($r',\phi'$) in relation to the receiving fiber coordinate system. On the other hand, the same point has coordinates ($r,\phi$) in relation to the optical spot coordinate system. The angle $\phi$ is defined positive in the clockwise direction with origin at the $y$ axis. Using the law of cosines on the triangle $Qoo'$, one can find the radial coordinate $r'$ written in terms of the coordinates $r$ and $\phi$:
\begin{equation}
r'^2 = r^2 + y_o^2 - 2ry_o \cos\phi,
\label{eq:rl}
\end{equation}
where $y_o$ is the distance from the center of the receiving fiber to the center of the optical spot. The variable $y_o$ is a function of $\theta$, since $y_o = m - Y_2(\theta)$, so:
\begin{equation}
y_o(\theta) = m - K\theta,
\label{eq:yo}
\end{equation}
where $m = b + b_R + \delta$, constant, is the distance between the center of the fibers. 
For a point located at the boundary of the receiving fiber core, $r'=a_R$, one can find the equation for the displaced circle in relation to the $r \phi$ system:
\begin{equation}
\phi(r) = \cos^{-1}\left(\frac{r^2 + y_o^2 - a_R^2}{2ry_o}\right).
\label{eq:phi}
\end{equation}	
Equation \ref{eq:Po} can be evaluated integrating twice the semi-circle on the region where $x > 0$ with integration limits for the variable $\phi$ ranging from $0$ to $\phi(r)$ and for the variable $r$, from $0$ to $y_o + a_R$. Thus, Eq. \ref{eq:Po} becomes:
\begin{equation}
P_o = \Gamma \frac{4 P_i}{\pi w^2} \int_{0}^{y_o+a_R}{\int_{0}^{\phi(r)}{r \exp\left(\frac{-2 r^2}{w^2}\right)}{\rm d}\phi}{\rm d}r.
\label{eq:Po2}
\end{equation}
Dividing Eq. \ref{eq:Po2} by $P_i$ and evaluating the integral on $\phi$, we obtain the power transfer coefficient as a function of $\theta$:
\begin{multline}
\eta(\theta) = \frac{P_o}{P_i} = \Gamma \frac{4}{\pi w^2} \int_{0}^{y_o(\theta)+a_R}{r \exp\left(\frac{-2 r^2}{w^2}\right)} \\ \times {\cos^{-1}\left(\frac{r^2 + y_o^2(\theta) - a_R^2}{2ry_o(\theta)}\right)}{\rm d}r.
\label{eq:eta}
\end{multline} 
For the particular case where $y_o=0$, i.e., the optical spot is centered on the receiving fiber core, Eq. \ref{eq:yo} yields $\theta=\theta_o=m/K$. At this point the function $\phi(r)$ is not defined and the value for $\eta(\theta_o)$ can be found integrating $\phi$ in Eq. \ref{eq:Po2} from $0$ to $\pi$, which results in:
\begin{multline}
\eta(\theta_o) = \Gamma \frac{4}{w^2} \int_{0}^{a_R}{r \exp\left(\frac{-2 r^2}{w^2}\right)} {\rm d}r = \\ \Gamma \left[1-\exp\left(\frac{-2 a_R^2}{w^2}\right)\right].
\label{eq:eta0}
\end{multline}

Finally, the static characteristic curve of the sensor can then be obtained evaluating the integral on Eq. \ref{eq:eta} numerically for each corresponding value of $\theta$, and using the result of Eq. \ref{eq:eta0} for $\theta=\theta_o$.
As can be seen, the reflective surface must have a non zero angle $\theta$, for the light to reach the receiving fiber core, since emitting and receiving fiber are spatially separated. However, in order to yield a static characteristic curve centered on zero, the axis of $\theta$ can be shifted by subtracting $\theta_o$.

\section{Results} 
Three optical fibers with different core radius were used in this work: one single-mode fiber (4/62.5~$\mu$m core/cladding radii) and two multimode fibers (25/62.5~$\mu$m and 52.5/62.5~$\mu$m core/cladding radii). The single-mode and multimode fiber numerical apertures were $NA = 0.12$ and $NA = 0.22$, respectively. Using these fibers, nine different sensor configurations were considered using one fiber as the emitting and the other as the receiving. The name given to the configuration was composed by the emitting fiber core radius ($a$) and the receiving fiber core radius ($a_R$), both in $\mu$m, as $a/a_R$. As an example, a configuration with $a=4$ $\mu$m and $a_R=25$ $\mu$m is called sensor 4/25.

The static characteristic curves simulated and experimentally acquired, were for the following sensors: 4/4, 4/25, 4/52.5, 25/4, 25/25, 25/52.5, 52.5/4, 52.5/25, and 52.5/52.5. The cladding radii of both emitting ($b$) and receiving ($b_R$) fibers, for all sensor configurations, were kept the same and unchanged: external diameter of 125 $\mu$m, i.e., $b = b_R = 62.5$ $\mu$m. Regarding an ideal condition, there is no gap between fibers ($\delta = 0$) and there are no transmission losses ($\Gamma = 1$). The focal length of the positive lens was $f_L = 7.4$ mm. The spot size over the receiving fiber plane was regarded constant as $w = a + 2$ $\mu$m. Regarding $w_o \approx 0.5$ mm and $\lambda = 532$ nm, results in $z_R \approx 1.5$ m, which means that $Z_1$ can be up to approximately 750 mm ($z_R/2$), as stated in the mathematical model section. The sensor was mounted (and simulated) with $Z_1$ much smaller than this value; it was chosen as $Z_1 = 35.4$ mm.

\subsection{Static characteristic curve simulation}
\begin{figure}[ht]
  \centering    
  \includegraphics[width=7cm]{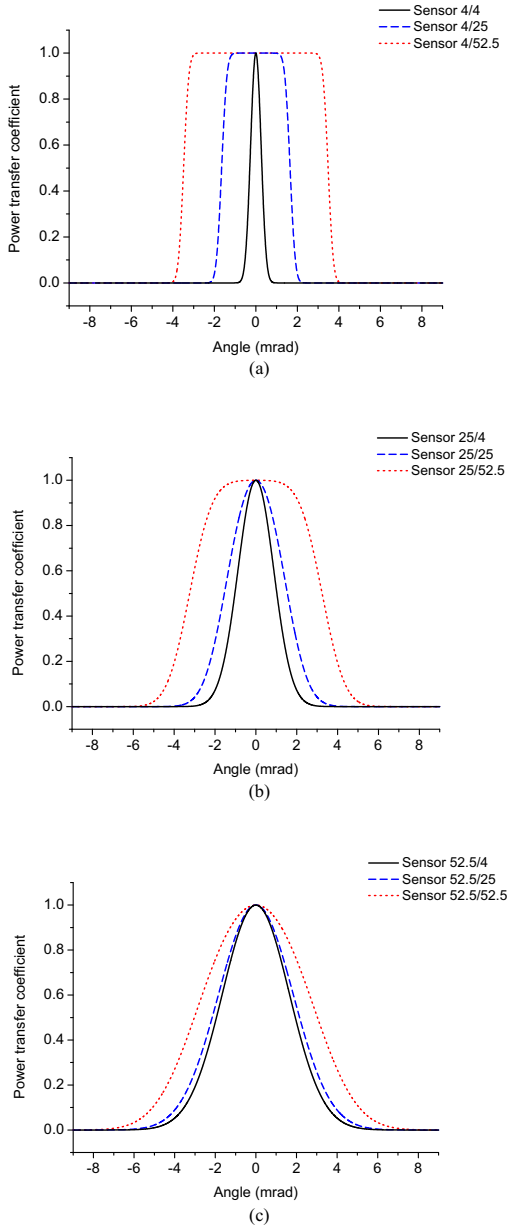}
  \caption{(Color online) Simulated static characteristic curves for the sensors: (a) 4/4, 4/25, and 4/52.5, (b) 25/4, 25/25, and 25/52.5, (c) 52.5/4, 52.5/25, and 52.5/52.5.}
  \label{fig:simul}
\end{figure}

The static characteristic curves were simulated with the software Matlab, using the parameters of the actual optical fibers and components. The first set of simulations was accomplished with the emitting fiber of 4 $\mu$m in three configurations: 4/4, 4/25, 4/52.5. The power transfer coefficient, $\eta$, was normalized and evaluated as a function of $\theta$. The simulation results are shown on Fig.~\ref{fig:simul} (a), where the solid line is the result for the sensor 4/4, the dashed line is the result for the sensor 4/25, and the dotted line is the result for the sensor 4/52.5.

In the second set of simulations, the core radius of the emitting fiber was changed to 25 $\mu$m. The receiving fiber core radius as in the former case was simulated with three different values: 4 $\mu$m, 25 $\mu$m and 52.5 $\mu$m, called respectively the sensor 25/4, 25/25 and 25/52.5. The simulation results are shown in Fig. \ref{fig:simul} (b), where the solid line is the result for the sensor 25/4, the dashed line is the result for the sensor 25/25, and the dotted line is the result for the sensor 25/52.5.

The third set of simulations was accomplished using an emitting fiber with a core radius of~52.5 $\mu$m. Once again, three curves were achieved for receiving fibers with core radius of: 4~$\mu$m, 25~$\mu$m and 52.5~$\mu$m. The sensors were called sensor 52.5/4, 52.5/25 and 52.5/52.5, respectively. The results are shown in Fig. \ref{fig:simul} (c), where the solid line is the result for the sensor 52.5/4, the dashed line is the result for the sensor 52.5/25, and the dotted line is the result for the sensor 52.5/52.5.

For all sensors' configurations, the curve is symmetric and there are two linear regions, one positive slope and the other, negative. It can be observed that, for the sensors which $a_R \leq a$, there is peak and for the sensors which $a_R > a$, the peak is extended to a flat region, as expected, since the optical spot is entirely confined on the receiving fiber core.  

Each static characteristic curve was normalized by its maximum output value, that corresponds to the peak, called in this text as $V_{\rm max}$. The side of the positive slope was chosen to fit a linear curve (using the least squares method) which linear range meets a criterion of nonlinearity less than or equal to 1\%. The operation point (or bias point) of the sensor is defined as the $\eta$ value corresponding to the center of the linear range and the normalized sensitivity is the inclination of the fitted curve. For each sensor, the normalized sensitivity, linear range, operation point, and nonlinearity were obtained, shown in Table~1.
\begingroup
\squeezetable
\begin{table}[h]
{\bf \caption{Simulation results$^a$}}
\begin{center}
\begin{tabular}{p{1.5cm}p{2.2cm}p{1.2cm}p{2cm}p{1cm}}
\hline
	Configu-ration
	& Normalized sensitivity \footnotesize{[mV/$\mu$rad]}
  & Linear range \footnotesize{[$\mu$rad]}
  & Operation point \footnotesize{[mV]}
  & Nonlin-earity \footnotesize{[\%]} \\
\hline
4/4	& $(2.40\times V_{\rm max})$ &	194 &	$(0.58\times V_{\rm max})$ & 0.98 \\
4/25 &	$(1.93\times V_{\rm max})$	& 223 & $(0.50\times V_{\rm max})$ & 1.00	\\
4/52.5 &	$(1.94\times V_{\rm max})$	& 221 & $(0.50\times V_{\rm max})$ & 0.98 \\
25/4	& $(0.67\times V_{\rm max})$	& 679 & $(0.60\times V_{\rm max})$ & 0.99 \\
25/25	& $(0.49\times V_{\rm max})$	& 949 & $(0.55\times V_{\rm max})$ & 0.99 \\
25/52.5 &	$(0.44\times V_{\rm max})$	& 979 & $(0.50\times V_{\rm max})$ & 1.00 \\
52.5/4	& $(0.36\times V_{\rm max})$	& 1276 & $(0.60\times V_{\rm max})$ & 0.98 \\
52.5/25	& $(0.32\times V_{\rm max})$	& 1424 & $(0.59\times V_{\rm max})$ & 0.99 \\
52.5/52.5	& $(0.25\times V_{\rm max})$	& 1840 & $(0.54\times V_{\rm max})$ & 1.00 \\
\hline
\end{tabular}
\end{center}
\footnotesize $^a$The parameter $V_{\rm max}$ must be in units of volts.
\end{table}
\endgroup

The normalized sensitivity is a merit factor that can be used to compare sensors' configurations each other. Analyzing the results for a given emitting fiber radius, the sensors with the smaller receiving fiber core radius have the higher sensitivity. Among these sensors, the one that also has the smaller emitting fiber core radius has the highest sensitivity, i.e., the sensor 4/4 is the most sensitive, $(2.40\times V_{\rm max})$~mV/$\mu$rad, with a linear range of 194~$\mu$rad. The sensor with the largest emitting and receiving core radii, i.e., the sensor 52.5/52.5 presented the smallest sensitivity, $(0.25\times V_{\rm max})$~mV/$\mu$rad, with a linear range of 1840~$\mu$rad.

A practical and direct calibration method can be used to determine the actual sensitivity (unnormalized) of the sensor, only by measuring $V_{\rm max}$ (in units of volts) and substituting the value on the corresponding normalized sensitivity.

\subsection{Experimental static characteristic curve}
\begin{figure}[ht]
  \centering
  \psfrag*{theta}[l]{\small $\theta$}
  \includegraphics[width=8.2 cm]{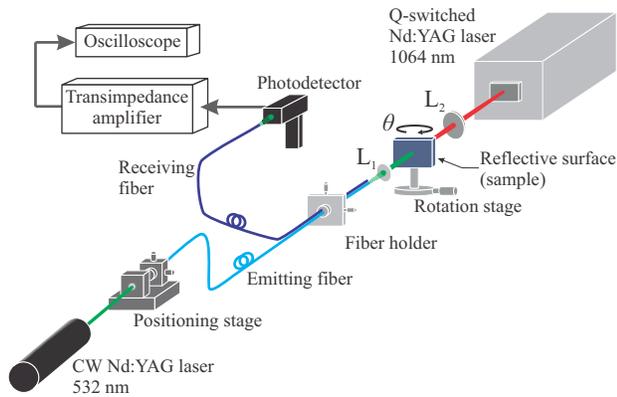}
  \caption{(Color online) Experimental setup.}
  \label{fig:set}
\end{figure}

In order to experimentally acquire the static characteristic curve for each sensor configuration and to verify the agreement with the simulation, an experimental setup was mounted as shown in Fig.~\ref{fig:set} (without the lens L$_2$ and the Q-switched laser).
The optical source used was a continuous wave (CW) Nd:YAG laser with wavelength of 532~nm and 55~mW of maximum output power. The reflective surface used for this experiment was a rigid mirror since we are interested in the static characteristic curve. The mirror was mounted over a rotation stage, driven by a micrometer, that was used to vary the angle $\theta$ and the corresponding output voltage was measured on the oscilloscope. When using the 4~$\mu$m core radius fiber (as emitting fiber), regardless of the fact that it is single-mode for 1310~nm and we used a 532~nm laser, the extra propagation modes were cut off by a mode filter which allowed only the propagation of the fundamental mode, LP$_{01}$. In this way, the optical spot delivered by the emitting fiber was symmetric, circular and homogeneous (without speckle). 

The assembly and adjustment of the sensor head showed to be very simple since it does not require displacement or angular alignment between fibers. The fibers (emitting and receiving) were placed parallel and glued together (using cyanoacrilate ester) with the aid of a regular microscope (amplification of 10$\times$) just to align their tips and to avoid gap between them (in order to ensure $\delta = 0$). The adjustment of the lens L$_1$ position was accomplished by projecting the light spot over a screen on different distances and verifying for the collimation. After that, the mirror was placed in front of L$_1$ and it was aligned such that the sensor provides the maximum output. The light collected by the receiving fiber was directed to a photodetector, part of a transimpedance amplifier circuit, which converts the input photocurrent to output voltage. This circuit was designed to detect frequencies from DC up to 85~MHz. An oscilloscope was used to acquire and record the voltage signal from the transimpedance amplifier. Care was taken to align the returning optical spot (reflected by the reflective surface) in the $x$ direction (remained fixed after alignment) to make sure that the spot translation on the $y$ direction was aligned on the diameter of the receiving fiber. Misalignments on the $x$ direction could provide a characteristic curve shrunken in relation to the actual curve.

\begin{figure}[ht]
\centering
	\includegraphics[width=7cm]{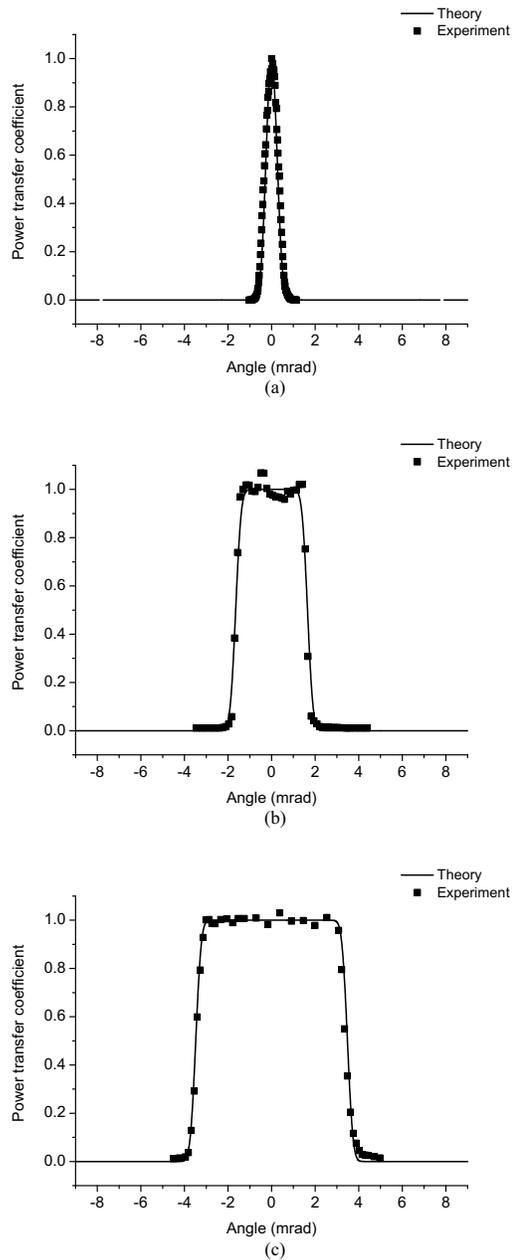}
  \caption{Experimental results: (a) Sensor 4/4, (b) Sensor 4/25, (c) Sensor 4/52.5.}
  \label{fig:res1}
\end{figure}

\begin{figure}[ht]
\centering
	\includegraphics[width=7cm]{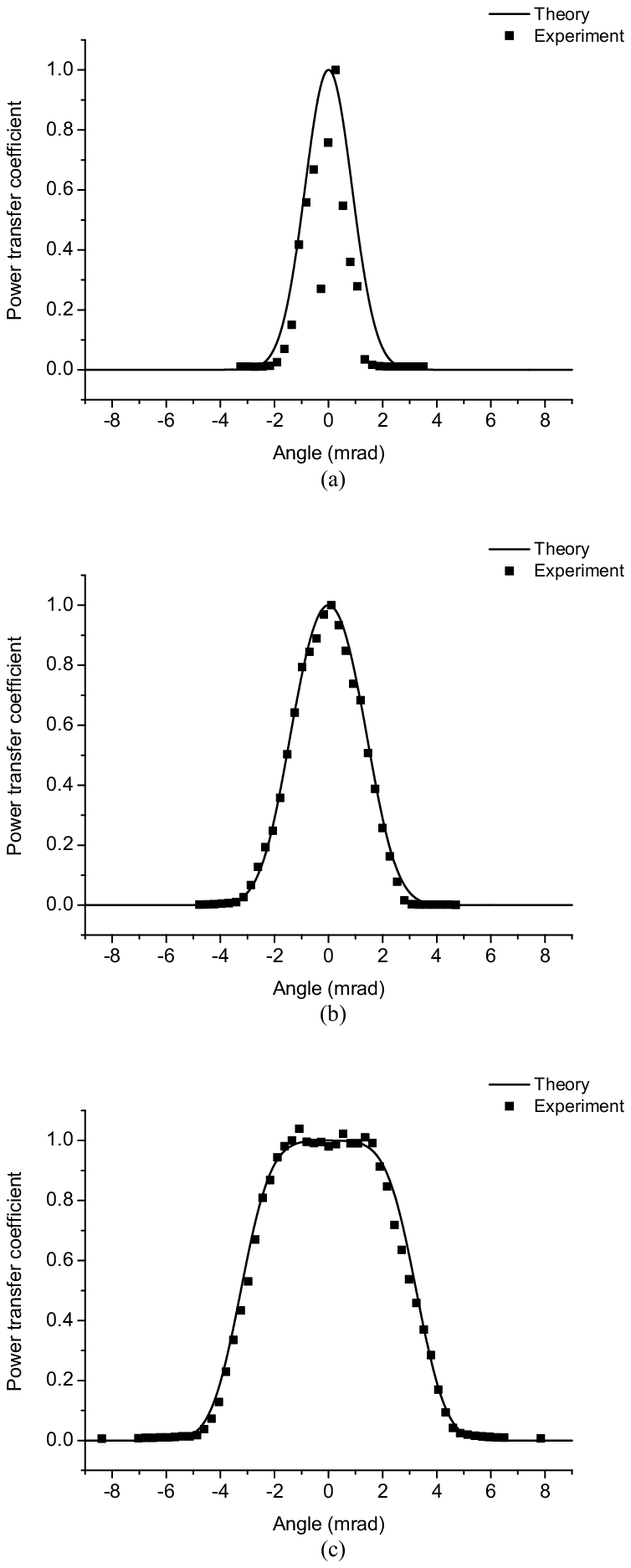}
  \caption{Experimental results: (a) Sensor 25/4, (b) Sensor 25/25, (c) Sensor 25/52.5.}
  \label{fig:res2}
\end{figure}

\begin{figure}[ht]
\centering
	\includegraphics[width=7cm]{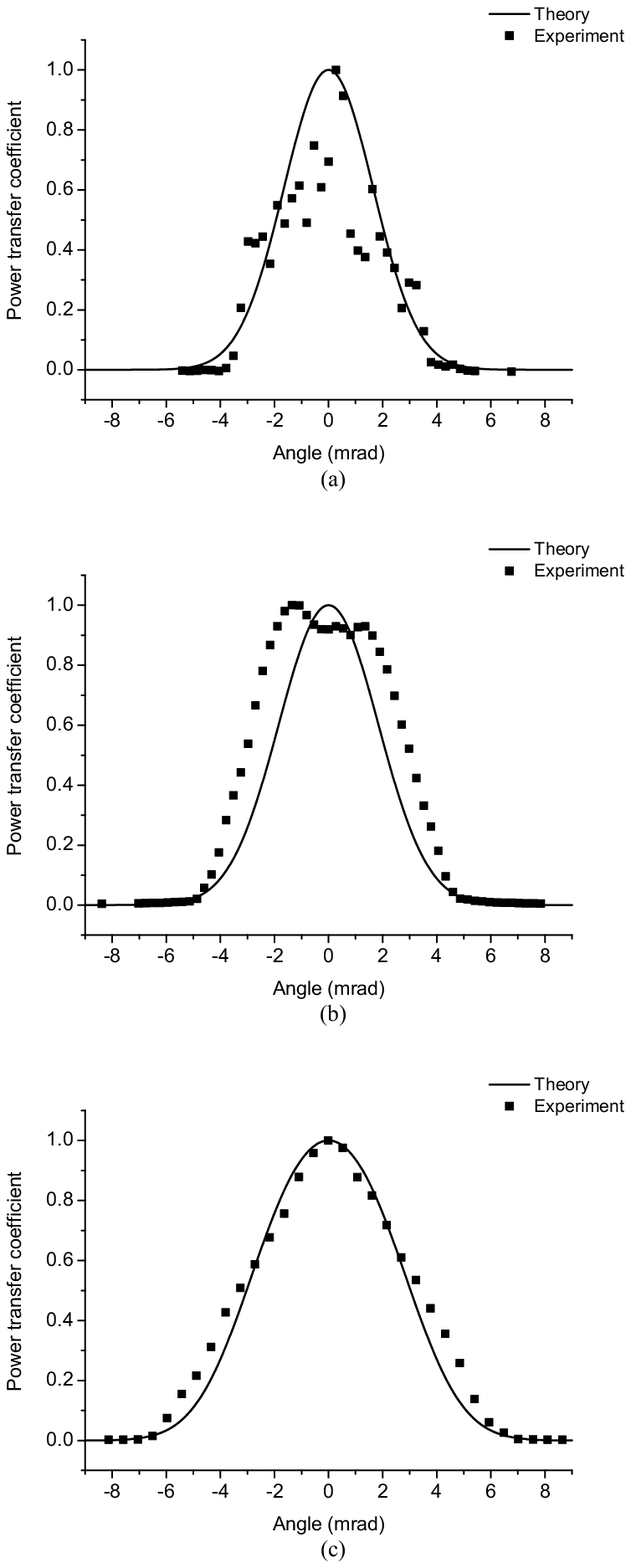}
  \caption{Experimental results: (a) Sensor 52.5/4, (b) Sensor 52.5/25, (c) Sensor 52.5/52.5.}
  \label{fig:res3}
\end{figure}

The experimental results are shown in Figs.~\ref{fig:res1}, \ref{fig:res2}, and \ref{fig:res3} in which the data were normalized for comparison with the mathematical model simulation. The experimental data is shown as square markers and the simulation is shown as a solid line. 
The mathematical model was corroborated by the experiment for the sensors 4/4, 4/25, 4/52.5, 25/25, 25/52.5, and 52.5/52.5 while the data of the sensors 25/4, 52.5/4 and 52.5/25 did not agree with the theory. The sensors 25/4 and 52.5/4 presented experimental data oscillating instead of the smooth behavior predicted by simulation. The experiment showed that the receiving fiber with $a_R$ = 4 $\mu$m works as a slit when the emitting fiber has $a = 25$ or 52.5 $\mu$m and the characteristic curve amplitude varies randomly as the speckles reach the receiving fiber core radius. In this way, the use of the receiving fiber with $a_R$ = 4 $\mu$m is recommended only when the emitting fiber has $a = 4$ $\mu$m, i.e., the 4/4 sensor. The sensor 52.5/25 presented a larger experimental curve and two peaks, probably because of the pattern of the modes coupled in the multimode emitting fiber. However, the sensor 52.5/25 can still be used for angular displacement measurements since its positive or negative slope keeps smooth and linear. The discrepancy between simulation and experiment for the sensors 25/4, 52.5/4 and 52.5/25 showed that care must be taken when using a multimode emitting fiber, which speckle pattern do not meet the homogeneous gaussian intensity profile assumed on the model; this can be a relevant issue when the receiving fiber core radius is smaller than the emitting fiber core radius. Nevertheless, the mathematical model is useful for designing the sensor configuration according to the application requirements (e.g., sensitivity and linear range). In addition, this model is general and can be used to test for variations in other parameters, e.g., cladding radius or the lens focal distance.

In order to obtain the highest sensitivity for this sensor, it was chosen the sensor configuration 4/4 which has the highest normalized sensitivity: $(2.40\times V_{\rm max})$~mV/$\mu$rad. Measuring the peak voltage with the rigid mirror (reflective surface with reflectivity of 100\%) it was obtained $V_{\rm max}=3.2$~V (DC voltage), which results in a sensitivity of 7.7~mV/$\mu$rad, for this specific setup.

\subsection{Ultrasound detection}
For dynamic operation, the fiber optic sensor configuration shall be chosen according to the sensitivity and linear range requirements of the measurement. In this case, the configuration used in all experiments was the sensor 4/4 due to its higher sensitivity. The sensor's reflective surface is then substituted by the sample under analysis and the alignment of the sensor head is accomplished. The $\theta$ angle is varied to find and measure $V_{\rm max}$ (DC voltage). Then, the operation point is set adjusting $\theta$ to a value which provides an output voltage of approximately $(0.58\times V_{\rm max})$ ($V_{\rm max}/2$ can also be used as a practical value). At this point, the sensor is ready to perform dynamic measurements and the oscilloscope can be set to AC acquisition.

\begin{figure}[ht]
  \centering
  \includegraphics[width=8.2 cm]{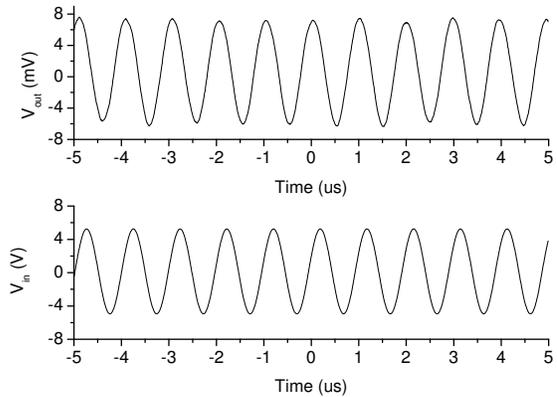}
  \caption{Output signal (top) and input signal (bottom) for the piezoelectric transducer.}
  \label{fig:piezo}
\end{figure}

\begin{figure}[ht]
  \centering
  \includegraphics[width=8.2 cm]{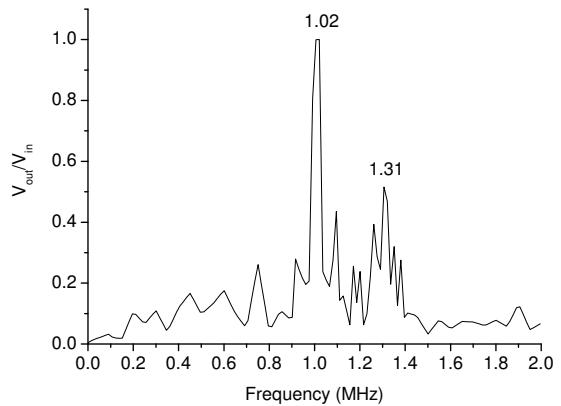}
  \caption{Frequency response of the piezoelectric transducer measured with the 4/4 sensor.}
  \label{fig:freqresp}
\end{figure}

\begin{figure}[ht]
  \centering
  \includegraphics[width=8.2 cm]{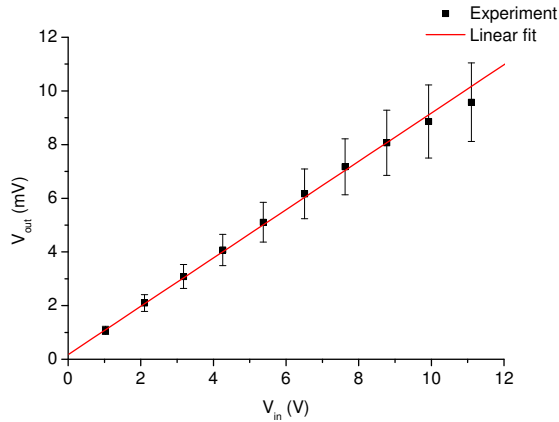}
  \caption{(Color online) Linearity of the piezoelectric transducer at 1.02 MHz.}
  \label{fig:linear}
\end{figure}

In the first experiment, a well-behaved signal (sinusoidal) was used to test the dynamic operation of the sensor. The experimental setup shown in Fig.~\ref{fig:set} was modified, substituting the reflective surface by a piezoelectric transducer (without the lens L$_2$ and the Q-switched laser). The optical source used was again the CW Nd:YAG laser (wavelength of 532~nm and 55~mW of maximum output power). The transducer resonance is centered on approximately 1~MHZ, according to the manufacturer. The transducer tip is circular with diameter of 35~mm; its surface is metallic (providing a reasonable optical reflectivity) leading to $V_{\rm max}=0.34$~V and, consequently, a sensitivity of 0.82~mV/$\mu$rad. A sinusoidal input voltage $V_{\rm in}=10.2$~V (peak-to-peak) with frequency of 1.02~MHz was applied to the transducer and the output voltage provided by the sensor was measured as $V_{\rm out}=14$~mV (peak-to-peak). The input and output voltage waveforms were acquired using the oscilloscope and they are shown in Fig.~\ref{fig:piezo}.
The frequency response of the transducer was then acquired varying the input frequency from 0.2 to 2~MHz and measuring the ratio between $V_{\rm out}$ and $V_{\rm in}$. The frequency response curve is shown in Fig.~\ref{fig:freqresp}, where the ratio $V_{\rm out}/V_{\rm in}$ was normalized.
Figure~\ref{fig:freqresp} shows the highest resonance amplitude at 1.02 MHz and a smaller resonance at 1.31 MHz. The linearity curve for the transducer was then acquired keeping the frequency on the resonance, i.e., 1.02MHz, and the voltage of $V_{\rm in}$ was varied from approximately 1 to 11 V (peak-to-peak). The result is shown in Fig.~\ref{fig:linear}. As can be seen on the graphic, the piezoelectric transducer presents a linear response over the range analyzed. 
These results show that the sensor is suitable to detect a sinusoidal ultrasonic wave in time domain and to measure the frequency response and linearity of a transducer.

A second experiment was performed aiming the detection of non sinusoidal signals: pulsed ultrasonic waves. For this, the laser ultrasonics experimental setup shown in Fig.~\ref{fig:set} was used. A Q-switched Nd:YAG laser, called generation laser, was used to generate ultrasonic pulses in aluminum samples. This type of laser can generate the longitudinal, shear and surface waves \cite{ledbetter1979, edwards1989, klein2009}. The Q-switched laser (wavelength of 1064 nm) was operated at single shot with energy of 420 mJ and a pulsewidth of 9 ns in all subsequent experiments. The laser spot diameter was approximately 6 mm.

\begin{figure}[ht]
  \centering
  \includegraphics[width=8.2 cm]{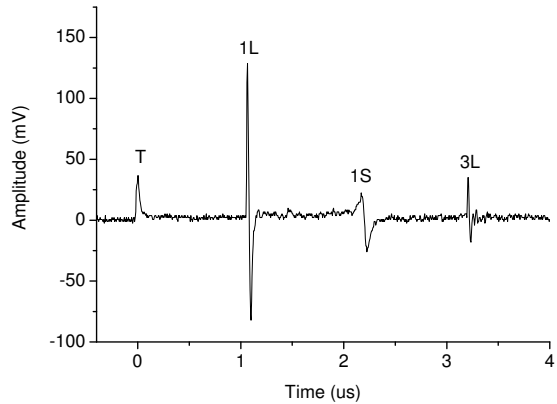}
  \caption{Longitudinal and shear wave detection on aluminum sample. T is the trigger, 1L is the first longitudinal wave arrival (after traveling through the thickness once), 3L is the second longitudinal wave arrival (after traveling the thickness three times), and 1S is the first shear wave arrival.}
  \label{fig:waveform1}
\end{figure}

The first laser ultrasonics measurement was accomplished with an aluminum sample with dimensions of $86 \times 30 \times 6.88$~mm, i.e., 6.88~mm thickness. This sample was polished on one side (to increase the amount of light reflected) and the other side was painted black (to increase light absorption). In this way, the generation laser was pointed to the black side of the sample and the sensor head, pointed to the polished surface. The generation laser spot and the sensor spot were aligned, in the sample thickness direction, for the detection of ultrasonic waves very close to their epicenters. Adjusting the angle $\theta$ of the sample, the maximum voltage output was measured as $V_{\rm max}=0.36$~V, which means a sensitivity of 0.86~mV/$\mu$rad. The lens L$_2$ was used to decrease the generation laser spot to approximately 3~mm at the sample surface. The fluence was 5.9~J/cm$^2$. The light pulse from the generation laser was used as a trigger on the oscilloscope to avoid jitter and delay from the electronic circuits.  The waveform acquired and registered in an oscilloscope is shown in Fig.~\ref{fig:waveform1}, where T corresponds to the trigger (light pulse), 1L corresponds to the first arrival of the longitudinal wave (after traveling through the thickness once) and 3L corresponds to the second arrival of the longitudinal wave (after traveling the thickness three times).
The distance traveled by the longitudinal wave between 1L and 3L corresponds to twice the thickness, i.e., $\Delta S_L = 2 \times 6.88$~mm. The arrival time of 1L and 3L were measured as $t_{1L}=1.065$~$\mu$s and $t_{3L}=3.205$~$\mu$s, respectively. Time delay between 1L and 3L was calculated as $t_L = t_{3L} - t_{1L} = 2.14$~$\mu$s and the longitudinal wave velocity was evaluated as $v_L = \Delta S_L/t_L = 6.43$~mm/$\mu$s. The shear wave arrival can be also observed in Fig.~\ref{fig:waveform1} as 1S. This wave traveled the thickness once, so $\Delta S_S = 6.88$~mm.  The shear wave arrival time was measured as $t_{1S}=2.17$~$\mu$s. The shear wave velocity was then evaluated as $v_S = \Delta S_S/t_S = 3.17$~mm/$\mu$s.
The ultrasonic velocity of the longitudinal wave on aluminum, according to \cite{ledbetter1979}, is approximately 6.42 mm/$\mu$s, which means an error of 0.2\%. For the shear wave, the velocity is approximately 3.13~mm/$\mu$s, meaning an error of 1.3\%.

\begin{figure}[ht]
  \centering  
  \includegraphics[width=8.2 cm]{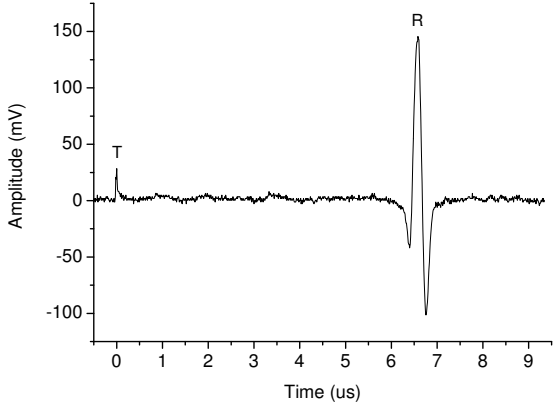}
  \caption{Rayleigh wave detection on aluminum sample. T is the trigger and R is the Rayleigh wave arrival.}
  \label{fig:waveform3}
\end{figure}

The second laser ultrasonics measurement was accomplished to detect a surface (Rayleigh) wave in an aluminum block with dimensions of $50 \times 50 \times 34$~mm. The setup shown in Fig.~\ref{fig:set} was modified with the generation laser pointing on the front surface of the sample, i.e., the polished side (the same as the sensor head). The maximum output voltage measured at the detection point was $V_{\rm max}=0.7$~V leading to a sensitivity of 1.68~mV/$\mu$rad. The generation laser was then focused with a positive cylindrical lens (instead of the spherical lens L$_2$) with 50~mm of focal length to generate a laser line with approximately 6~mm of height on the sample surface. The distance between the generation laser line and the sensor spot was $\Delta S_R = 19.5$~mm. The resulting signal was acquired with the oscilloscope and is shown in Fig.~\ref{fig:waveform3}.
In this figure, T is the trigger pulse and R is the Rayleigh wave arrival. The time delay between the trigger and the Rayleigh arrival measured was $t_R = 6.58$~$\mu$s. Then, the resulting Rayleigh velocity was $v_R = \Delta S_R/t_R = 2.96$~mm/$\mu$s. 
According to \cite{ledbetter1979}, the velocity of the Rayleigh wave in aluminum is approximately 2.93~mm/$\mu$s, which means an error of 1\%.

\section{Conclusions}
The fiber optic sensor developed in this work is highly sensitive to angular displacement, has a linear response, wide bandwidth, and is simple and low cost. Compared with other fiber optic sensors, a bundle of fibers is not necessary, the fibers can be positioned further from the reflective surface, the assembly and adjustment of parameters is simpler, and it uses few optical components. Regarding a hypothetical version using only one fiber (to both emit and receive the light), the advantage of using two fibers is the simplicity of the setup and the cost, since it would require additional expensive optical components. Besides, the sensor provides an output voltage signal in the time domain directly, with no additional circuits for demodulation as occurs in the interferometer configurations.

The simulation results showed that the sensor configurations tested are highly sensitive to small angle variation (in the range of microradians) with nonlinearity less than or equal to 1\%. Moreover, the simulation showed that the fiber core radius (both the emitting and the receiving fiber) influences the normalized sensitivity of the sensor: it is inversely proportional to the fibers core radius. As a result, the sensor 4/4 presented the highest normalized sensitivity, $(2.40\times V_{\rm max})$~mV/$\mu$rad (with a linear range of 194~$\mu$rad), and the sensor 52.5/52.5 presented the lowest, ten times smaller, $(0.25\times V_{\rm max})$~mV/$\mu$rad (with a linear range of 1840~$\mu$rad). It was observed that the actual sensitivity (unnormalized) of the sensor depends on the reflectivity of the reflective surface. Using a mirror as a reflective surface, the sensor 4/4 presented a sensitivity of 7.7~mV/$\mu$rad.

The sensors 4/4, 4/25, 4/52.5, 25/25, 25/52.5, and 52.5/52.5 showed a good agreement between the simulation and experiment, while the sensors 25/4, 52.5/4 and 52.5/25 did not agree. For the case of the sensors 25/4 and 52.5/4, the disagreement is due to the combination of the speckle pattern from the multimode emitting fibers (25 and 52.5~$\mu$m core radius) and the relatively small core radius of the receiving fiber which is 4~$\mu$m for both sensors. The receiving fiber in this case works as a slit and, as a consequence, the intensity can increase or decrease according to the speckle position. The sensor 52.5/25 provided a result in disagreement with the theory however, it can still be used for ultrasonic measurements since its positive or negative slope stays smooth and linear. The mathematical model showed to be general and useful for designing the sensor configuration according to the application requirements.

The sensor was initially tested to detect the sinusoidal vibration of a piezoelectric actuator. The time domain waveform, the frequency response and the linearity of the actuator were acquired and showed the potential of the sensor for the detection of ultrasonic waves and for characterization of piezoelectric actuators. As part of a laser ultrasonics system, the sensor was able to detected longitudinal, shear, and surface waves (generated by the Q-switched laser). The velocity of the longitudinal, shear, and surface waves were measured on aluminum samples as 6.43~mm/$\mu$s, 3.17~mm/$\mu$s, 2.96~mm/$\mu$s, respectively, and the presented error was smaller than 1.3\%. The sensor proved to be suitable for time-of-flight measurements and nondestructive inspection, being an alternative to the piezoelectric or the interferometric detectors. 

\section{Acknowledgments}
The authors wish to thank Dr. Marcelo G. Destro for providing the multimode fibers and the cylindrical lens used in this work, Dr. Julio C. Adamowski for providing the piezoelectric transducer, Dr. Rog{\'e}rio M. Cazo for the support on the design and manufacturing of the electronic circuits, Dr. Nicolau A. S. Rodrigues and Dr. Rudimar Riva for the discussions about the optical problems. One of the authors (JMSS) acknowledges the Brazilian sponsor agencies Conselho Nacional de Desenvolvimento Cient{\'i}fico e Tecnol{\'o}gico (CNPq), for the provision of a scholarship, and the Coordena{\c c}{\~a}o de Aperfei{\c c}oamento de Pessoal de N{\'i}vel Superior (CAPES), for the provision of an international scholarship.


\begin{thebibliography}{99}
\bibitem{scruby1990a}{C. B. Scruby and L. E. Drain, ``Introduction,'' in \emph{Laser ultrasonics: techniques and applications}, (Adam Hilger, 1990), pp. 1--36.}
\bibitem{murfin2000}{A. S. Murfin, R. A. J. Soden, D. Hatrick, and R. J. Dewhurst, ``Laser-ultrasound detection systems: a comparative study with Rayleigh waves,'' Meas. Sci. Technol. \textbf{11}, 1208--1219 (2000).}
\bibitem{monchalin2002}{J. -P. Monochalin, C. N{\'e}ron, M. Choquet, A. Blouin, B. Reid, D. L{\'e}vesque, P. Bouchard, C. Padioleau, and R. H{\'e}on, ``Detection of flaws in materials by laser-ultrasonics,'' in \emph{IUTAM Symposium on Advanced Optical Methods and Applications in Solid Mechanics}, A. Lagarde, ed. (Springer Netherlands, 2002), pp. 437--450.}
\bibitem{sorazu2003}{B. Sorazu, G. Thursby, B. Culshaw, F. Dong, S. G. Pierce, Y. Yang, and D. Betz, ``Optical generation and detection of ultrasound,'' Strain \textbf{39}, 111--114 (2003).}
\bibitem{monchalin1986}{J. -P. Monchalin, ``Optical detection of ultrasound,'' IEEE Transactions on Ultrasonics, Ferroelectrics and Frequency Control \textbf{33}, 485--499 (1986).}
\bibitem{monchalin2004}{J. -P. Monochalin,``Laser-ultrasonics: from the laboratory to industry,'' in \emph{AIP Conference Proceedings}, D. O. Thompson, D. E. Chimenti, L. Poore, C. Nessa, and S. Kallsen, eds. (AIP, 2004), pp. 3--31.}
\bibitem{bramhavar2009}{S. Bramhavar, B. Pouet, and T. W. Murray, ``Superheterodyne detection of laser generated acoustic waves,'' \apl \textbf{94}, (2009).}
\bibitem{beard1997}{P. C. Beard and T. N. Mills, ``Miniature optical fibre ultrasonic hydrophone using a Fabry-Perot polymer film interferometer,'' Electron. Lett. \textbf{33}, (1997).}
\bibitem{ling2011}{T. Ling, S. -L. Chen, and L. J. Guo, ``High-sensitivity and wide-directivity ultrasound detection using high Q polymer microring resonators,'' \apl \textbf{98}, (2011).}
\bibitem{dewhurst1999}{R. J. Dewhurst and Q. Shan, ``Optical remote measurement of ultrasound,'' Meas. Sci. Technol. \textbf{10}, R139֭R168 (1999).}
\bibitem{perret2011}{L. Perret, L. Chassagne, S. Top{\c c}u, P. Ruaux, B. Cagneau, and Y. Alayli, ``Fiber optics sensor for sub-nanometric displacement and wide bandwidth systems,'' Sens. Actuators A \textbf{165}, 189--193 (2011).}
\bibitem{menadier1967}{C. Menadier, C. Kissinger, and H. Adkins, ``The fotonic sensor,'' Instrum. Cont. Syst. \textbf{40}, 114--120 (1967).}
\bibitem{he1991}{G. He and F. W. Cuomo, ``A light intensity function suitable for multimode fiber-optic sensors,'' J. Lightwave Technol. \textbf{9}, 545--551 (1991). }
\bibitem{ko1995}{W. H. Ko , K.-M. Chang, and G.-J. Hwang, ``A fiber-optic reflective displacement micrometer,'' Sens. Actuators A \textbf{49}, 51֭55 (1995).}
\bibitem{shimamoto1996}{A. Shimamoto and K. Tanaka, ``Geometrical analysis of an optical fiber bundle displacement sensor,'' \ao \textbf{35}, 6767--6774 (1996).}
\bibitem{bergougnoux1998}{L. Bergougnoux, J. Misguich-Ripault, and J. L. Firpo, ``Characterization of an optical fiber bundle sensor,'' Rev. Sci. Instrum. \textbf{69}, 1985--1990 (1998).}
\bibitem{faria1998}{J. B. Faria, ``A theoretical analysis of the bifurcated fiber bundle displacement sensor,'' IEEE Transactions on Instrumentation and Measurement \textbf{47}, 742--747 (1999).}
\bibitem{zheng1999}{J. Zheng and S. Albin, ``Self-referenced reflective intensity modulated fiber optic displacement sensor,'' Opt. Eng. \textbf{38}, 227--232 (1999).}
\bibitem{bucaro2001}{J. A. Bucaro, N. Lagakos, ``Lightweight fiber optic microphones and accelerometers,'' Rev. Sci. Instrum. \textbf{72}, 2816--2821 (2001).}
\bibitem{buchade2006}{P. B. Buchade and A. D. Shaligram, ``Simulation and experimental studies of inclined two fiber displacement sensor,'' Sens. Actuators A \textbf{128}, 312--316 (2006).}
\bibitem{buchade2007}{P. B. Buchade and A. D. Shaligram, ``Influence of fiber geometry on the performance of two-fiber displacement sensor,'' Sens. Actuators A \textbf{136}, 199--204 (2007).}
\bibitem{patil2011}{S. S. Patil and A. D. Shaligram, ``Modeling and experimental studies on retro-reflective fiber optic micro-displacement sensor with variable geometrical properties,'' Sens. Actuators A \textbf{172}, 428--433 (2011).}
\bibitem{moro2011}{E. A. Moro, M. D. Todd, and A. D. Puckett, ``Using a validated transmission model for the optimization of bundled fiber optic displacement sensors,'' \ao \textbf{50}, 6526--6535 (2011).}
\bibitem{sagrario1998}{D. Sagrario and P. Mead, ``Axial and angular displacement fiber-optic sensor,'' \ao \textbf{37}, 6748--6754 (1998).}
\bibitem{bois1989}{E. Bois, S. J. Huard, and G. Boisde, ``Loss compensated fiber-optic displacement sensor including a lens,'' \ao \textbf{28}, 419--420 (1989).}
\bibitem{wu1995}{C. Wu, ``Fiber optic angular displacement sensor,'' Rev. Sci. Instrum. \textbf{66}, 3672--3675 (1995).}
\bibitem{khiat2010}{A. Khiat, F. Lamarque, C. Prelle, N. Bencheikh, and E. Dupont, ``High-resolution fibre-optic sensor for angular displacement measurements,'' Meas. Sci. technol. \textbf{21}, 1--10 (2010).}
\bibitem{wang1997}{H. Wang, ``Collimated beam fiber optic position sensor: effects of sample rotations on modulation functions,'' Opt. Eng. \textbf{36}, 8--14 (1997).}
\bibitem{feldmann2005}{M. Feldmann and S. Buttgenbach, ``Microoptical distance sensor with integrated microoptics applied to an optical microphone,'' in \emph{Sensors, 2005 IEEE}, (IEEE, 2005), pp. 769--771.}
\bibitem{sakamoto2010}{J. M. S. Sakamoto and G. M. Pacheco,``Theory and experiment for single lens fiber optical microphone,'' Physics Procedia \textbf{3}, 651--658.}
\bibitem{siegman1971}{A. E. Siegman, ``Optical resonators and lens waveguides,'' in \emph{An Introduction to Lasers and Masers},
 (McGraw-Hill, 1971), pp. 293--345.}
\bibitem{siegman1971a}{A. E. Siegman, ``Physical properties of gaussian beams,'' in \emph{An Introduction to Lasers and Masers},
 (McGraw-Hill, 1971), pp. 663--697.}
\bibitem{ledbetter1979}{H. M. Ledbetter and J. C. Moulder, ``Laser-induced Rayleigh waves in aluminum,'' J. Acoust. Soc. Am. \textbf{65}, 840--842 (1979).}
\bibitem{edwards1989}{C. Edwards, G. S. Taylor, and S. B. Palmer, ``Ultrasonic generation with a pulsed TEA CO$_2$ laser,'' J. Phys. D: Appl. Phys. \textbf{22}, 1266--1270 (1989).}
\bibitem{klein2009}{M. B. Klein and H. Ansari, ``Signal processing techniques for nondestructive evaluation using laser ultrasonics,'' in Proceedings of IEEE International Symposium on Signal Processing and Information Technology (IEEE, 2009), p. 316.}

\end{thebibliography}
\end{document}